\documentclass{llncs}
\usepackage{graphicx}
\usepackage{dcolumn}
\newcolumntype{d}[1]{D{.}{.}{#1}}
\begin{document}

\title{The Digital Flynn Effect: Complexity of Posts on Social Media Increases over Time}
\titlerunning{The Digital Flynn Effect}  % abbreviated title (for running head)
\author{Ivan Smirnov}
\authorrunning{Ivan Smirnov} % abbreviated author list (for running head)
\tocauthor{Ivan Smirnov}
\institute{National Research University Higher School of Economics, Moscow, Russia\\
\email{ibsmirnov@hse.ru}
}

\maketitle              % typeset the title of the contribution

\begin{abstract}
Parents and teachers often express concern about the extensive use of social media by youngsters. Some of them see emoticons, undecipherable initialisms and loose grammar typical for social media as evidence of language degradation. In this paper, we use a simple measure of text complexity to investigate how the complexity of public posts
on a popular social networking site changes over time. We analyze a unique dataset that contains texts posted by $942,336$ users from a large European city across nine years. We show that the chosen complexity measure is correlated with the academic performance of users: users from high-performing schools produce more complex texts than users from low-performing schools. We also find that complexity of posts increases with age. Finally, we demonstrate that overall language complexity of posts on the social networking site is constantly increasing. We call this phenomenon the digital Flynn effect. Our results may suggest that the worries about language degradation are not warranted.
\end{abstract}
\section{Introduction}
Media reports often express concern about the extensive use of social media by young people \cite{drouin2011college}. One major concern is that texting style typical for social media leads to poor language skills. However, no clear links between texting practices and literacy were discovered (see \cite{wood2013text,zebroff2017youth} for a review of the topic). As social media are constantly evolving, it might also be important to account for potential changes in language practices of its users. Does the complexity of texts posted on social media decrease, indicating the language degradation that parents and educators are afraid of? We address this question by using a large longitudinal dataset.

We investigate the changes in complexity of public posts made by users on a popular social networking site across nine years. We use data from VK, the largest European social networking site, that provides functionality similar to Facebook. The dataset contains $1,320,572,032$ words posted by $942, 336$ users from Saint-Petersburg, Russia whose ages range from $15$ to $60$. In their profiles, users indicate the high school in which they study or which they graduated from. In addition to post content, information about average academic performance of graduates from these schools is also available (see Methods for details).

We use average word length as the measure of text complexity. The traditional measures of text complexity are readability indices that are based on average word length, average sentence length or number of complex words in the text \cite{gunning1952technique,flesch1979write}. The same features were successfully used for single-sentence readability prediction in the Russian language  \cite{karpov2014single}. In the online context, average word length was shown to be shorter for the simple English Wikipedia than for the main English Wikipedia \cite{yasseri2012practical}. In this paper, we limit ourselves to average word length (see Methods for details). It is worth mentioning that automatic determination of sentence length is more problematic in the social media context than in traditional texts. For example, people often neglect to capitalize the first letter of a sentence, and emoticons can be used to indicate the end of a sentence but may appear in the middle of a sentence as well.

Measuring text complexity is a challenging task. One might even argue that using shorter words means that authors have found a clever way to communicate the same amount of content more efficiently. In order to justify our measuring system, we compare average word length in posts made by users with different academic performance. We expect that users with higher academic performance produce more complex posts. We also expect to see age-related differences in the posts’ complexity. It was previously reported that the average word length in texts increases with the age of their authors \cite{nguyen2013old,pennebaker2003words}.

\begin{figure}[hb!]
\centering
\includegraphics[scale=0.5]{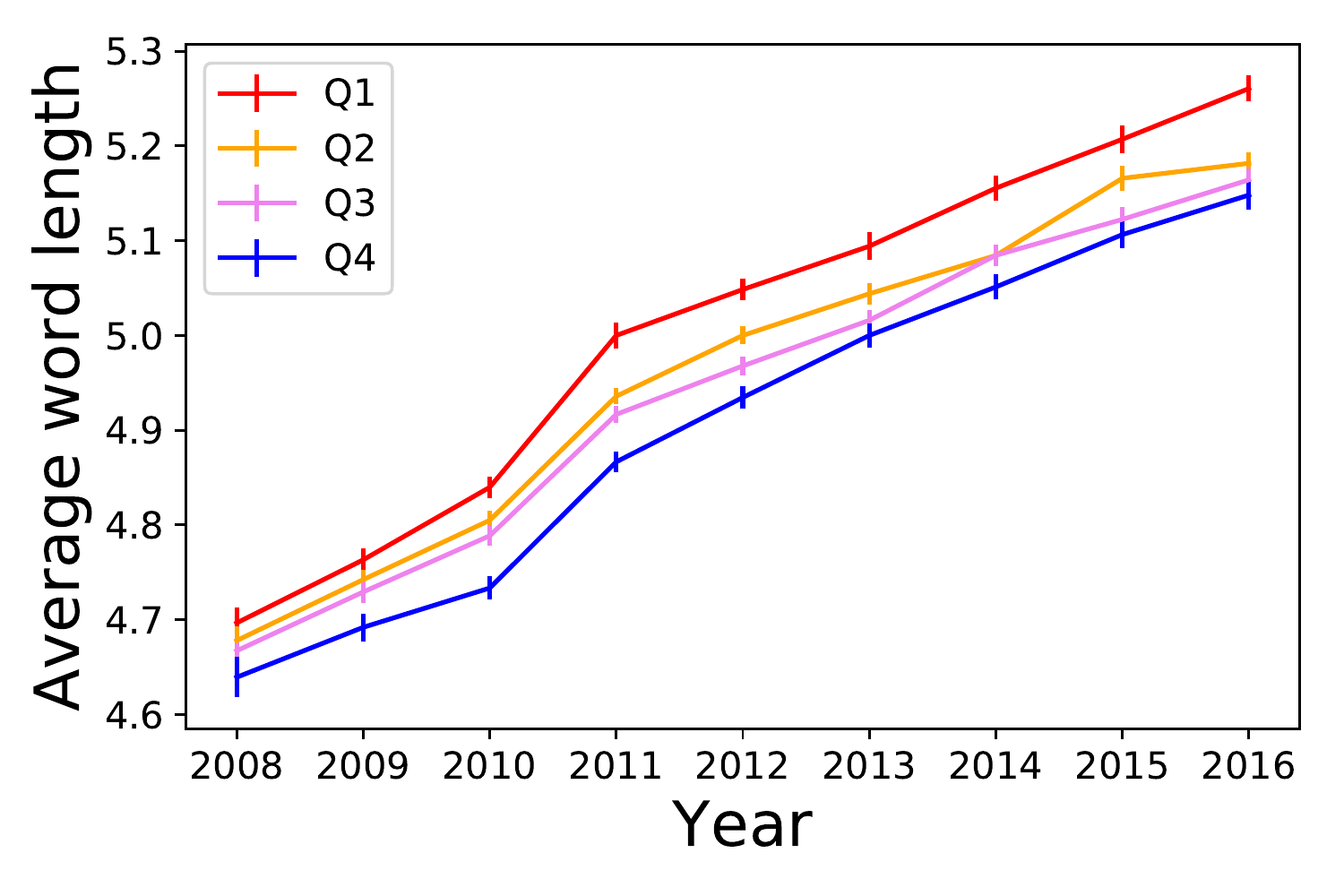}
\caption{\textbf{Average word length in posts of graduates from different schools.} Users
from top-performing schools (Q1) produce more complex posts than users from middle-
or low-performing schools. However, there is no significant difference between middle-high (Q2) and middle-low (Q3) performing schools. Vertical bars are standard errors.}
\label{Fig:performance}
\end{figure}

\section{Results}
For each school in our dataset, we assign a quartile ranking based on the average academic performance of its graduates. Q1 denotes the top-performing schools and Q4 the low-performing schools. We find that the complexity of posts is correlated with users’ academic performance (Fig. \ref{Fig:performance}): users from high-performing schools (Q1) produce more complex texts than those from middle-performing schools (Q2, Q3), and users from middle-performing schools produce more complex texts than those from low-performing schools (Q4). However, we find no significant difference between middle-high (Q2) and middle-low (Q3) performing schools.

\begin{figure}[ht!]
\centering
\includegraphics[scale=0.5]{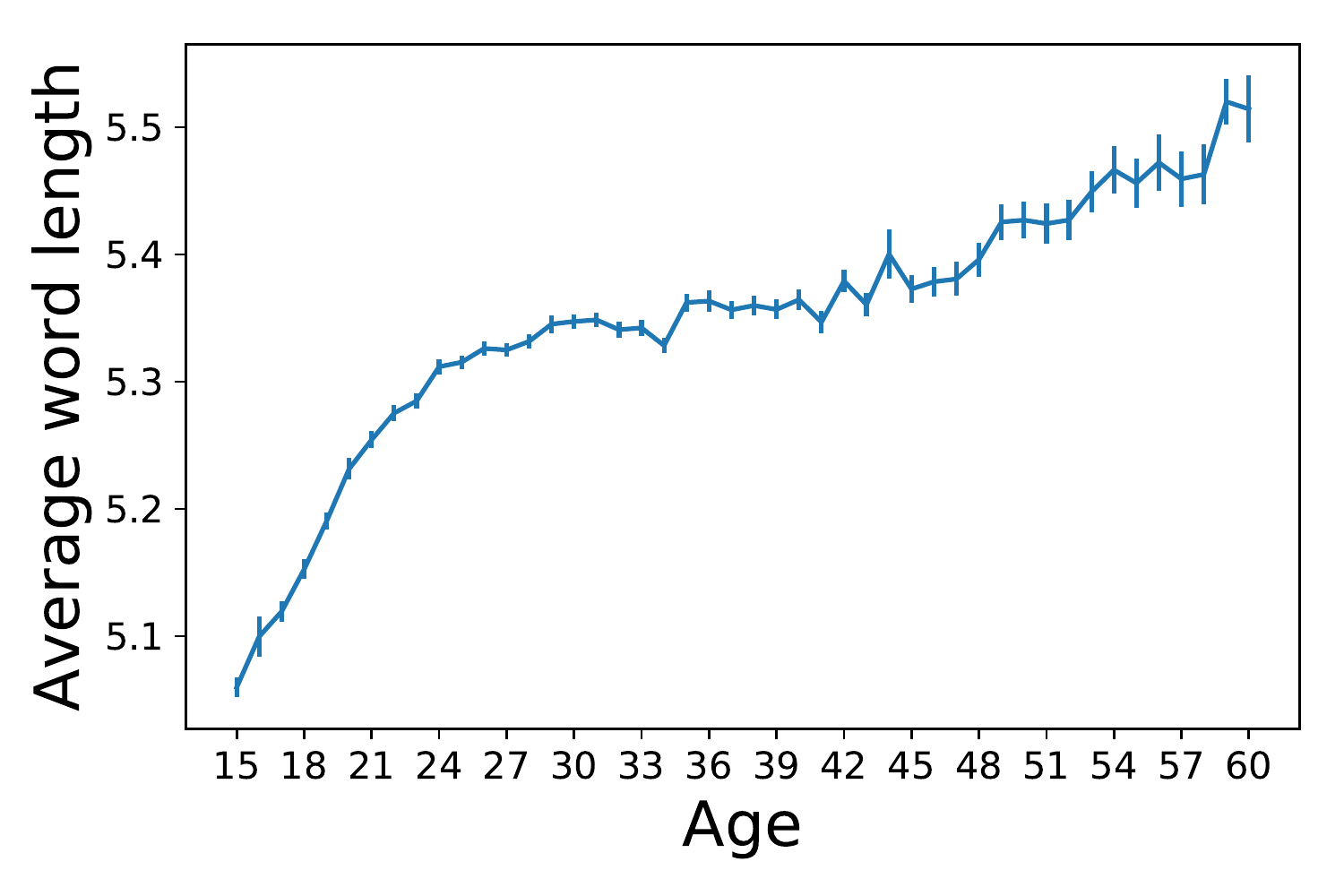}
\caption{\textbf{Average word length in posts made by users of different ages in 2016.} The average word length increases with age until the late 20s. The increase after the early 40s is potentially explained by the sample bias. Vertical bars are standard errors.}
\label{Fig:age}
\end{figure}

We find that the complexity of posts increases with age until the late 20s (Fig. \ref{Fig:age}). It is relatively stable through the 30s and increases again starting from the early 40s. The latter fact is potentially explained by the sample bias. While VK use is ubiquitous among young people, the proportion of older people who use the social networking site is significantly lower. Less than half of users are in their 40s ($124,830$) compared to those in their 30s ($287,641$) in the dataset.

\begin{figure}[ht!]
\centering
\includegraphics[scale=0.5]{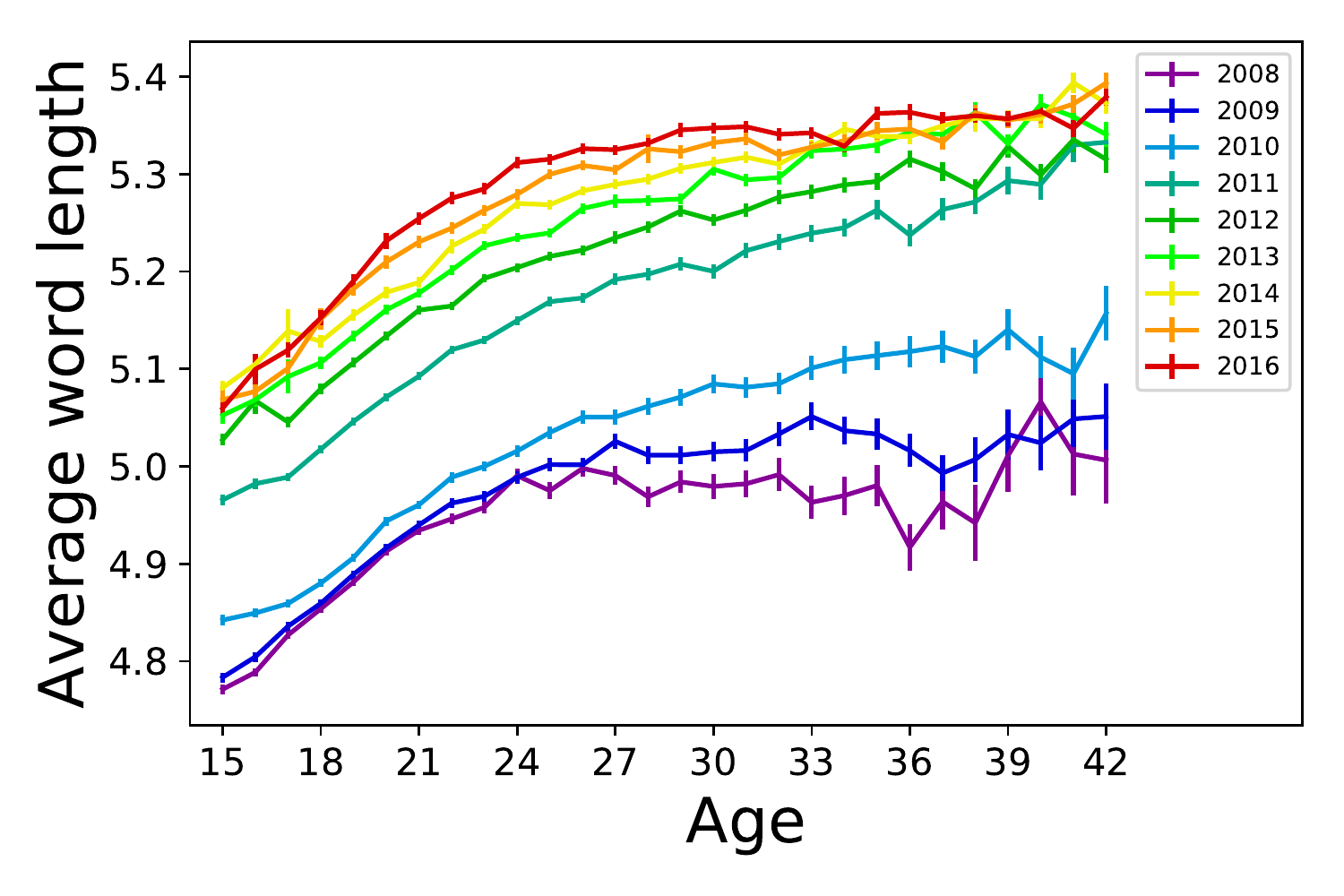}
\caption{\textbf{The digital Flynn effect.} The complexity of posts on a social networking site increases over time. Each line represents posts made by users of different ages in a given year. Vertical bars are standard errors.}
\label{Fig:time}
\end{figure}

Finally, we compare the complexity of posts at different time points. We find that complexity increases over time (Fig. \ref{Fig:time}). This increase cannot be explained by aging alone: 15-year-old users in 2016 wrote more complex posts than users of any age in 2008. We call this phenomenon the digital Flynn effect by analogy with the so-called Flynn effect --- the massive increase in IQ test scores over the course of the twentieth century \cite{flynn1984mean,flynn1987massive,flynn2007intelligence}.

\begin{figure}[ht!]
\centering
\includegraphics[scale=0.5]{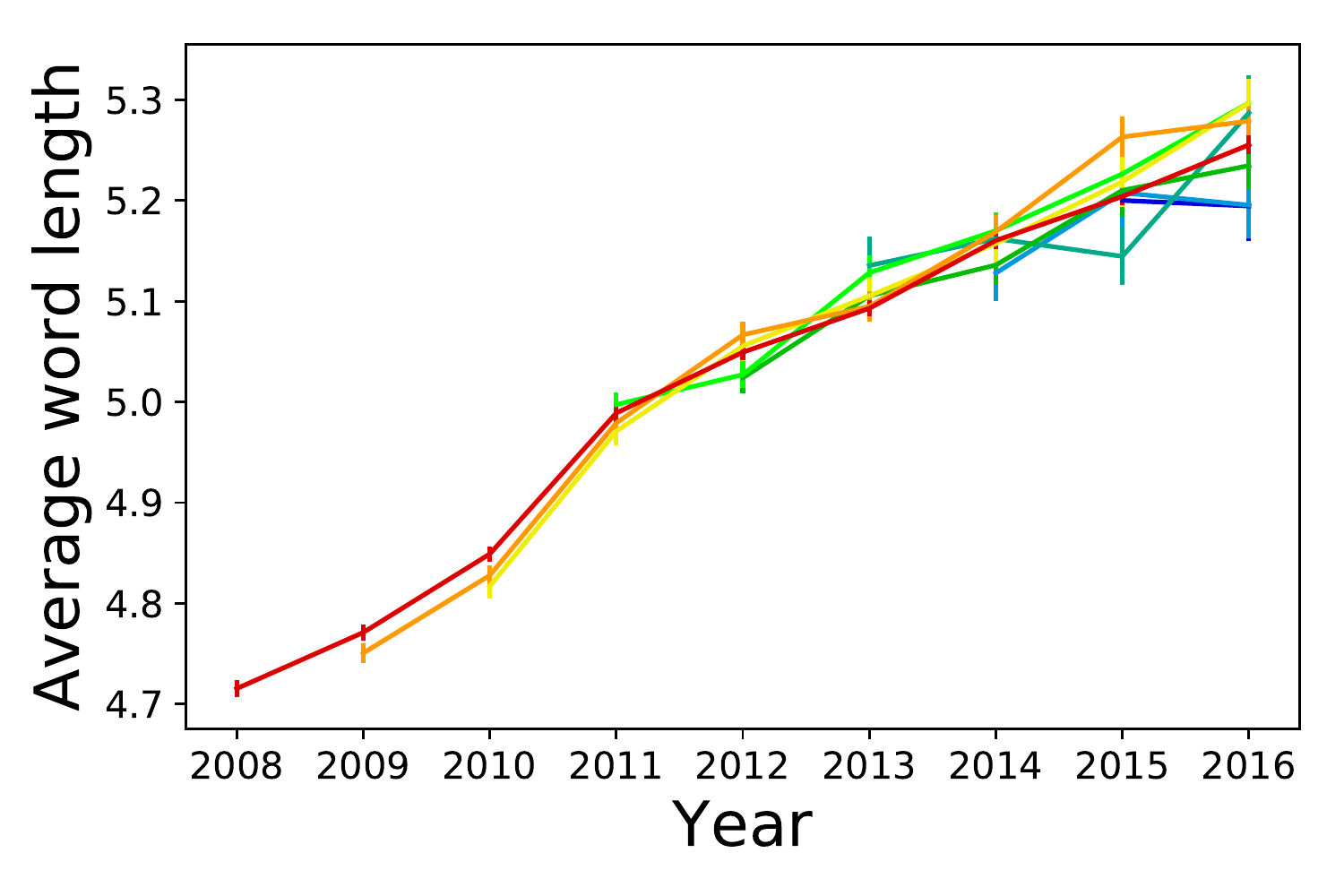}
\caption{\textbf{Time of site adoption and post complexity.} Each line corresponds to users who started to post on the social networking site in a given year. The observed growth in text complexity cannot be explained by time of site adoption by different groups of users. Vertical bars are standard errors.}
\label{Fig:adoption}
\end{figure}

Intriguingly, the observed growth cannot be explained by differences in characteristics of users who joined the social networking site at different time points. The average word length follows the same line, regardless of the year when the first post on VK was made (see Fig. \ref{Fig:adoption}).

These results are summarised in Table \ref{Tbl:regression}. A multiple linear regression was calculated to predict the average word lengths of posts made by a group of users based on their age (0 corresponds to 15 years old), the time since their first post in years and the time passed since the launch of VK in years (0 corresponds to 2008 in both cases). The regression was calculated for 15- to 23-year-old users, the period when we observe linear growth in post complexity.

\begin{table}
\caption{\textbf{Coefficients from the regression model.} Both age and year since the launch of VK correlate with complexity, however time since site adoption does not.}
\begin{center}
\begin{tabular}{ld{4,6}r}
\hline
\multicolumn{1}{l}{Parameter} & \multicolumn{1}{l}{Estimate} & \multicolumn{1}{l}{Std. Err} \\
\hline
Intercept  &     4.734^{***} & 0.009\\
Age &   0.022^{***} & 0.001\\
Year since site adoption  & -0.002 & 0.002 \\
Year since launch of VK  & 0.050^{***} & 0.001\\
\hline
\multicolumn{2}{l}{$^{***}$ - p-value $< 10^{-3}$. $R^2 = 0.828$.}
\end{tabular}
\end{center}
\label{Tbl:regression}
\end{table}

The largest increase in text complexity occurs between 2010 and 2011. We track changes in complexity with higher time resolution and find that the largest gain in average word length was between October 2010 and November 2010. This is probably related to the major change in site functionality on the 20th of October of that year. On this day, the News Feed and Wall were consolidated (a similar change in functionality was made by Facebook in 2008).

\section{Methods}
VK is the largest European social networking site, with more than 100 million active users. It was launched in September 2006 in Russia and provides functionality similar to Facebook. According to VK’s Terms of Service: ``Publishing any content on his / her own personal page, including personal information, the User understands and accepts that this information may be available to other Internet users taking into account the architecture and functionality of the Site''.

VK provides an application programming interface (API) that enables downloading of information systematically from the site. In particular, it is possible to download user profiles from certain educational institutions and within selected age ranges. For each user, it is possible to obtain a list of their public posts. Posting times are known with a time resolution of one second. VK’s support team confirmed to us that the data downloaded via their API may be used for research purposes.

Using specially-developed software, the profiles of all users from Saint-Petersburg were downloaded. We excluded users who have no VK friends from the same educational institution that was indicated in their profile. It was previously shown that this is an effective way to remove fake user profiles \cite{smirnov2016search}.

High school graduates are obliged to pass the standardized examination (Unified State Examination or USE) in Russia. The information about average USE scores of graduates for each school in Saint-Petersburg is publicly available for certain years. We use the distribution of these scores to assign to each school a corresponding quartile, splitting them into four groups: high-performing schools, middle-high-performing schools, middle-low-performing schools, and low-performing schools. The results for one cohort of students (born in 1993) to which the data is available are shown in Fig. \ref{Fig:performance}. The relationship between time of site adoption and complexity growth in Fig. \ref{Fig:adoption} is shown for the same cohort. Similar results were obtained for other cohorts.

We define a word as a sequence of Cyrillic letters. Words preceded by the pound sign (\#) were excluded because hashtags typically contain multiple words without spaces between them. We also excluded words that contain the same letter three or more times in a row, in order to account for potential differences in word-lengthening practices among different groups of users. All posts written by a given user during a year were combined in one text, with the average length of words (mean value) in this text then being computed. We excluded users who wrote less than five posts in a given year. We also excluded all posts which contain a URL, in order to account for potential automatic posting by applications or websites. Only the original content written by users (not reposts of content written by someone else) were included in the dataset.

\section{Discussion}
We use average word length to measure language complexity of posts on a popular social networking site. We show that this measure is consistent with the assumed cognitive complexity of the posts: average word length increases with age and correlates with academic performance of users. We find that the complexity of posts is constantly increasing, and that this increase cannot be explained by aging alone. We call this phenomenon the digital Flynn effect.

Our results suggest that social media do not lead to language degradation. Instead, users change their language practices in this environment to produce more sophisticated texts than in previous years. It remains unclear whether the observed changes are specific to the particular social networking site, or reflect broader changes of language practices in online or even real-world settings. While there is not enough evidence to suggest a specific explanation of the observed changes, our results indicate that the way texts in social media are produced is constantly evolving. It might be important to investigate these changes in detail and to account for them in future research.

\bibliographystyle{splncs}
\bibliography{digital}

\end{document}